\documentclass[12pt]{article}
\usepackage{amsmath,amssymb,amsfonts,amsthm}
\usepackage{graphicx}
\usepackage{xcolor, colortbl}
\usepackage{empheq}
\usepackage[pdftex, bookmarks=true,colorlinks,linkcolor=red,urlcolor=blue, citecolor=blue]{hyperref}
\usepackage{cite}
\usepackage{subcaption}
\usepackage{slashed}

\newcommand{\etaL}{\eta_{L}}

\usepackage{tikz}
\usepackage{amsthm}
\usepackage{array}
\usepackage{soul}
\usepackage{float}
\usepackage{bm}

\makeatletter\@addtoreset{equation}{section}\makeatother

\renewcommand{\title}[1]{\vbox{\center\LARGE{#1}}\vspace{5mm}}
\renewcommand{\author}[1]{\vbox{\center#1}\vspace{5mm}}
\newcommand{\address}[1]{\vbox{\center\em#1}}

\newcommand{\be}{\begin{equation}}
\newcommand{\ee}{\end{equation}}
\newcommand{\bea}{\begin{eqnarray}}
\newcommand{\eea}{\end{eqnarray}}

\definecolor{darkgreen}{rgb}{0,0.3,0}
\definecolor{darkblue}{rgb}{0,0,0.3}
\definecolor{darkred}{rgb}{0.7,0,0}

\begin{document}

\unitlength=.8mm
\begin{titlepage}
\begin{center}
\hfill \\
\hfill \\
\vskip 1cm
\title{\boldmath Nonlinear nature of near-equilibrium 
viscous fluids}

\vskip 0.5cm
{Yan Liu$^{a, b}$}\footnote{Email: {\tt yanliu@buaa.edu.cn}} 
{and Hao-Tian Sun$^{c,b}$}\footnote{Email: {\tt sunhaotian@buaa.edu.cn}}

\address{$^{a}$Department of Space Science, Beihang University, Beijing 100191, China}

\address{$^{b}$Peng Huanwu Collaborative Center for Research and Education, \\ Beihang University, Beijing 100191, China}

\address{$^{c}$Crete Center for Theoretical Physics,\\ Institute for Theoretical and Computational Physics, Department of Physics,\\ University of Crete, Heraklion, Greece.}

\end{center}
\vskip 1.5cm
\abstract{
We study the late-time relaxation of a neutral relativistic viscous fluid in
$d+1$ dimensions. In the long-wavelength regime, linearized
hydrodynamics predicts that the sound mode at momentum $nk$ decays as
$e^{-n^2\omega_I t}$.  However, nonlinear analysis gives a decay of \(e^{-n\omega_I t}\).  
We derive a closed asymptotic attractor solution in which the frequency of the $n$-th
harmonic locks to $n$ times the complex frequency of the fundamental mode.  The amplitude envelopes for energy current $J$ obey a simple cascading relation,  $J_n=\alpha_J^{\,n-1}J_1^n$, with $\alpha_J$ fixed by the equation
of state, the longitudinal viscosity, and the {fundamental} wavenumber. For conformal fluids,
$\alpha_J=1/(8\eta k)$, in agreement with the holographic result of Ref.~\cite{Liu:2025hfv}.   
The existence of the attractor shows that, even near equilibrium, 
field powers are not equivalent to amplitude order. 
}

\end{titlepage}
	\begingroup 
	\hypersetup{linkcolor=black}
	\tableofcontents
	\endgroup

\section{Introduction}

Hydrodynamics  provides a universal effective description of many-body systems at
long wavelengths and late times.  For a neutral relativistic system, the slow
variables are the conserved energy and momentum densities.  Microscopic physics enters only through the
equation of state and transport coefficients, which makes relativistic hydrodynamics a common framework for the quark-gluon plasma, relativistic
astrophysical matter, strongly coupled quantum systems, and holographic black
branes \cite{Kovtun:2012rj,Romatschke:2009im}. 
In a dissipative system, a far-from-equilibrium state is expected to relax
towards global equilibrium. A basic question is then how this relaxation takes
place, and what is the precise dynamics near equilibrium.

Near global equilibrium, the standard treatment is to linearize the conservation
laws (and constitutive relations).  This reduces the nonlinear hydrodynamic
equations to an eigenvalue problem which gives the quasinormal modes of infinitesimal disturbances around
equilibrium.  At linear order, different Fourier modes evolve independently,
and each mode is characterized by its quasinormal frequency: its real part gives
oscillation, while its imaginary part gives damping. For sound,
\be
    \omega(k)=c_s k-i\Gamma_s k^2+\cdots , 
\ee
where \(c_s\) is the speed of sound and \(\Gamma_s\) is the sound attenuation
constant.
In
holography, these sound modes are realized as quasinormal modes of the dual black brane,
relating boundary relaxation to black-brane ringdown
\cite{Kokkotas:1999bd,Berti:2009kk,Kovtun:2005ev}.

This dispersion already contains an important hierarchy.  Modes with different
momenta do not relax at the same rate: for sound, the damping rate is
proportional to \(k^2\).  Thus higher Fourier modes are more strongly damped
than lower ones. As time evolves, different momentum sectors can
become automatically separated in amplitude which 
 calls into question the assumption that linearization is uniformly valid for all modes. 
In the full hydrodynamic equations,
the nonlinear sources built from
longer-lived lower modes can then become the dominant contribution in a higher-momentum sector.  Consequently,
the late-time behavior of that sector is not governed by its linear mode. This was the main finding of Ref.~\cite{Liu:2025hfv}: the system approaches
equilibrium along a unique late-time trajectory, which we call
an attractor. 

In this paper we analyze this mechanism in a neutral relativistic viscous fluid in \(d+1\)
dimensions. To simplify the calculation, we assume the system has one compact spatial direction, \(x\sim x+L\).  The compact
direction discretizes the allowed momenta into harmonics \(k_n=nk\), where $k={2\pi}/{L}$.  This setup also describes the
boundary fluid of the compactified black brane system studied in
Ref.~\cite{Liu:2025hfv}.

Our main result is a nonlinear late-time solution for energy density $E$ and energy current density $J$.  The \(n\)-th
harmonic does not relax with its intrinsic linear frequency. 
Instead its leading late-time
contribution  locks to \(n\) times the complex frequency of the fundamental mode 
\be \omega(k_n)\rightarrow n\, \omega(k).\ee
The corresponding late-time solution for the energy current $J$ is
\begin{align}
    J=\sum_{n\ge1}J_n e^{-n\omega_I t}\cos{(n\omega_Rt-\phi_n)}\sin{(nkx)}
\end{align}
The harmonic envelopes form a cascade,
\be
    J_n=\alpha_J^{\,n-1}J_1^n ,
\ee
where \(\alpha_J\) is fixed by the equation of state, the longitudinal
viscosity, and the fundamental wavenumber. The phase $\phi_n$ is fixed by the equation of state. For a conformal neutral fluid, \(\alpha_J=1/(8\eta k)\), in agreement with the holographic result of
Ref.~\cite{Liu:2025hfv}.

The resulting locked tower is what we call a nonlinear late-time hydrodynamic attractor.  At late times, the higher
harmonics are fixed by thermodynamics, transport and the lowest fundamental mode. 
In this sense, their memory of initial data is erased from the terminal solution. 

The paper is organized as follows.  Sec.~\ref{sec:linear} reviews the linear
sound spectrum and the late-time hierarchy among Fourier sectors.
Sec.~\ref{sec:n3} derives the nonlinearly forced solutions for the second, third, and general harmonics, and constructs the resulting harmonic cascade. 
Sec.~\ref{sec:cascade_conclusion} collects the attractor solution and its
limits.  Sec.~\ref{sec:discussion} discusses implications and limitations.

\section{Linear hydrodynamic spectrum and late-time hierarchy}
\label{sec:linear}
	
We begin by analyzing the linear hydrodynamic spectrum of a neutral viscous
fluid.   Starting from the first-order constitutive relation, we derive the
sound-mode dispersion relation and the momentum-dependent attenuation of the
Fourier harmonics.  We then use this damping hierarchy to show why linearized
evolution alone does not determine the late-time evolution in 
higher-momentum sectors. 

The equations of motion are the conservation equations for the energy-momentum
tensor\footnote{
We work in natural units \(c=1\) and use the mostly-plus metric
\(g_{\mu\nu}=(-,+,\ldots,+)\).
}
\begin{equation}
	\partial_\mu T^{\mu\nu}=0 .
	\label{eq:coneq}
\end{equation}
In the Landau frame, the constitutive relation up to first order in derivatives is\footnote{Higher derivative terms in the constitutive relation can be found, for example, in \cite{Baier:2007ix, Bhattacharyya:2007vjd, Grozdanov:2015kqa}; they are subleading at small momentum.} 
\begin{equation}
	T^{\mu\nu}
	= \epsilon u^\mu u^\nu + p\Delta^{\mu\nu}
	-\eta\sigma^{\mu\nu} -\zeta\Delta^{\mu\nu}\partial_\lambda u^\lambda
	+\mathcal{O}(\partial^2),
	\label{eq:T_neutral}
\end{equation}
where
\begin{equation}
	\Delta^{\mu\nu}=g^{\mu\nu}+u^\mu u^\nu,
	\qquad
	\sigma^{\mu\nu}
	= \Delta^{\mu\alpha}\Delta^{\nu\beta}
	\Bigl( \partial_\alpha u_\beta+ \partial_\beta u_\alpha
	-\frac{2}{d}\,\Delta_{\alpha\beta}\,\partial_\lambda u^\lambda \Bigr).
	\label{eq:sigma_def}
\end{equation}
Here \(\epsilon\) and \(p\) are the energy density and pressure in the comoving
frame, \(u^\mu\) is the four-velocity, and \(\eta\) and \(\zeta\) are the shear
and bulk viscosities.  The transverse projector satisfies
\(\Delta^{\mu}_{\ \mu}=d\).

We consider a static observer and introduce the conserved densities together with the longitudinal momentum flux
\begin{equation}
	\mathcal{E}\equiv T^{tt},\qquad J\equiv T^{tx},\qquad \Pi\equiv T^{xx}.
	\label{eq:EJ_def_neutral}
\end{equation}
Throughout the nonlinear analysis below we will also use the energy
perturbation measured by this static observer,
\begin{equation}
	E\equiv\mathcal{E}-\epsilon_0 .
	\label{eq:E_def_neutral}
\end{equation}
where $\epsilon_0$ is the energy density of global equilibrium. For our one-dimensional setup the conservation equations then reduce to
\begin{align}
	\partial_t \mathcal{E}+\partial_x J = 0, \qquad
	\partial_t J +\partial_x \Pi = 0.
	\label{momJ}
\end{align}
These equations become a closed hydrodynamic system once \(\Pi\) is expressed
in terms of $E$ and $J$ through the constitutive relation and the
equation of state.

\subsection{Linear expansion and dispersion relation}

Consider a state close to equilibrium with constant background quantities
\(\epsilon_0\), \(p_0\), and \(u^\mu_0=(1,0,\dots,0)\).  We introduce a
longitudinal perturbation depending only on \(t\) and \(x\):
\begin{equation}
	\epsilon=\epsilon_0+e,\qquad
	u^\mu=\gamma(1,v,0,\dots,0),\qquad
	\gamma=1/\sqrt{1-v^2},
	\label{eq:velocity_ansatz}
\end{equation}
with \(|e|\ll\epsilon_0\) and \(|v|\ll 1\).  To linear order in the
perturbations,
	\begin{align}
		\mathcal{E} 
        = \epsilon_0+e,~~~~~
		J 
        = w_0 v,~~~~~
		\Pi 
        = p_0+c_s^2 e - \etaL \partial_x v,
		\label{eq:Txx_linear}
	\end{align}
where
\begin{equation}
	p = p_0 + c_s^2 e + \cdots,\qquad
	c_s^2 = \Bigl(\frac{\partial p}{\partial\epsilon}\Bigr)_0.
	\label{eq:EOS_neutral_second_order}
\end{equation}
We have defined the equilibrium enthalpy \(w_0\equiv\epsilon_0+p_0\) and the
longitudinal viscosity
\(\etaL\equiv\zeta+\frac{2(d-1)}{d}\eta\).

It is useful to rewrite the constitutive relation directly in terms of the
conserved variables.  To linear order \(e=E=\mathcal{E}-\epsilon_0\) and
\(v=J/w_0\), so
\begin{equation}
	\Pi = p_0 + c_s^2E
	- \frac{\etaL}{w_0}\partial_x J.
	\label{eq:Txx_EJ_linear}
\end{equation}
Substituting this expression into the conservation equations gives the linear
hydrodynamic equations
\begin{align}
	\partial_t \mathcal{E}+\partial_xJ=0,\qquad
	\partial_tJ + c_s^2\partial_x\mathcal{E}
	- \frac{\etaL}{w_0}\partial_x^2J =0 .
	\label{eq:linear-hydro}
\end{align}

We impose periodic boundary conditions and decompose the perturbations into
Fourier modes,
\begin{align}
	E(t,x) = \sum_{n=1}^\infty e_n(t,x)=\sum_{n=1}^\infty e_n(t)\cos(nkx),
	\qquad
	J(t,x) =\sum_{n=1}^\infty j_n(t,x)= \sum_{n=1}^\infty j_n(t)\sin(nkx),
	\label{expand_J}
\end{align}
with \(k=2\pi/L\). 
Each Fourier sector evolves independently at linear order.
The energy equation gives the mode relation
\begin{equation}
	\dot e_n+nk j_n=0.
	\label{eq:emode_relation_EJ}
\end{equation}
For the momentum mode \(j_n(t)\), the equations reduce to
\begin{equation}
	\ddot j_n + c_s^2 n^2k^2 j_n
	+ \frac{\etaL}{w_0}n^2k^2 \dot j_n = 0.
	\label{eq:jn_linear}
\end{equation}

Thus the eigenfrequencies are
\begin{equation}
	\omega_n \equiv \omega_{nR} - i\omega_{nI}
	= c_s n k - i\frac{\etaL}{2w_0}(nk)^2
	+ \mathcal{O}\bigl((nk)^3\bigr).
	\label{eq:omega_n}
\end{equation}
This is the standard hydrodynamic sound mode dispersion relation \cite{Kovtun:2012rj, Romatschke:2009im, Hartnoll:2018xxg}.  In
particular, the long-wavelength regime for the fundamental  mode can be measured
by the damping ratio
\begin{equation}
	\varepsilon \equiv \frac{\omega_{I}}{\omega_{R}}
	= \frac{\etaL}{2w_0c_s}\,k \ll 1,
	\qquad
	\omega_R=c_s k,\qquad
	\omega_I=\frac{\etaL}{2w_0}k^2 ,
	\label{eq:epsilon_def}
\end{equation}
where $\omega_R$ and $\omega_I$ are the real frequency and damping rate of the fundamental mode, respectively.

\subsection{Linear truncation is not consistent for all momentum sectors
}
\label{sec:linear_inconsistency}

This linear spectrum implies an important point: if the equations of motion
contain nonlinear terms, then the linear truncation is not uniformly consistent
across all momentum sectors.

To see this, take initial data in different Fourier sectors to be of the same
order, and let the system evolve to a late time at which it can be described by quasinormal modes. 
Since the linear damping rate obeys
\(\omega_{nI}=n^2\omega_I\), one finds, up to prefactors determined by the initial data,
\begin{equation}
	j_n(t)\propto \left[j_1(t)\right]^{n^2}.
\end{equation} 
This means different momentum sectors are in different order of amplitude. For example, the equation for $2k$ momentum mode contains a nonlinear source proportional to $\left[j_1(t)\right]^2$. This source decays more  slowly than the intrinsic linear \(2k\) mode and eventually gives $j_2(t)\propto \left[j_1(t)\right]^2$, rather than $\left[j_1(t)\right]^4$. 
Therefore the nonlinear term must be
included in the equation of motion for the \(2k\) mode.

This comparison shows that {\em the power of fluctuation fields and the order of amplitude are not the same}. The condition $e\ll1$ does not by itself imply that $e^2\ll e$ is uniformly satisfied across momentum sectors. This observation motivates a joint power-momentum analysis:  one should keep all terms that can contribute in each harmonic sector and then analyze which terms dominate. Equations for different momenta can therefore require  different truncation power of nonlinear terms.

\subsubsection{Fixed point of the nonlinear decay spectrum}
We first give a rough estimate of the order of amplitude in different momentum sectors. Higher-momentum modes are generated by nonlinear terms constructed from slower-decaying lower modes. We assume
\begin{align}
    j_n(t)\gg j_{n+1}(t). 
    \label{eq:nonlinear spectrum}
\end{align}
This relation is satisfied by the linear modes, and for nonlinear equations, it can also be obtained by analyzing arbitrary products of modes in the spectrum of the system.

The equation of motion for the \(n\)-th harmonic contains quadratic and, in
general, higher-order products of other modes.  Therefore, before determining
the late-time behavior of a given momentum sector, one has to determine which
nonlinear source gives the slowest decay in that sector.

We start from the linear spectrum and focus on its imaginary part, which
controls the amplitude decay.  Since nonlinear terms couple different
momenta, it is useful to keep the decay exponent and the momentum label
separate.  For the linear sound modes,
\begin{align}
    \omega_{nI}=n^2\omega_I,\qquad
    k_n=nk .
\end{align}
Quadratic products of the \(n_a\)-th and \(n_b\)-th linear modes then generate
new candidate contributions with
\begin{align}
    \omega_{\{n_a,n_b\}I}
    =\omega_I(n_a^2+n_b^2),\qquad
    k_{\{n_a,n_b\}}=(n_a+n_b)k
    \quad \text{and} \quad
    k_{\{n_a,n_b\}}=|n_a-n_b|k .
\end{align}
Higher-order products generate analogous candidates: their decay exponents are
given by the sum of the decay exponents of the factors, while their momenta are
given by all allowed sums and differences of the factor momenta.

The first update is obtained by taking the union of the linear spectrum with
the spectra generated by all nonlinear products and, in each momentum sector,
keeping the smallest available decay exponent.  For example, in the \(2k\)
sector, the linear mode has decay exponent \(4\omega_I\), whereas the quadratic
product of two fundamental modes gives a contribution with decay exponent
\(2\omega_I\).  Thus the slowest available decay exponent in the \(2k\) sector
is updated from \(4\omega_I\) to \(2\omega_I\).

We then repeat the same construction using the updated spectrum: nonlinear
products of the updated modes generate a new set of candidate decay exponents,
which is again combined with the previous set, and the slowest exponent in each
momentum sector is retained.  Iterating this procedure gives a self-consistent
fixed point, namely a spectrum that is unchanged by further nonlinear products.
This fixed-point spectrum determines the leading late-time decay hierarchy,
\begin{align}
    \omega_{nI}=n\omega_I,\qquad
    k_n=nk .
\end{align}
This fixed point of the spectrum identifies the slowest late-time decay exponent among all possible nonlinear source terms available in each momentum sector at late times. Thus, at late time, it gives the ordering of $nk$ mode in  \eqref{eq:nonlinear spectrum}.

\subsubsection{Linear persistence of the fundamental mode }
The above analysis does not change the leading solution for the fundamental mode. The equation for the fundamental mode can still be truncated to the linear equation, because all nonlinear terms give only small late-time corrections.

To see this, projecting the conservation equations onto the first harmonic gives
\begin{equation}
	\ddot j_1 + c_s^2 k^2 j_1 + \frac{\etaL}{w_0} k^2 \dot j_1
	= \mathcal{S}_1[\{j_n\}],
	\label{eq:j1_full}
\end{equation}
where $\mathcal{S}_1$ denotes nonlinear terms with a $\sin(kx)$
projection.  A product of two fundamental modes produces only the 
zero momentum sector and the
second harmonics at momentum $2k$, so there is no quadratic feedback into $j_1$.  The first
possible feedback either involves higher harmonics 
(products of the type $j_mj_{m\pm 1}$ with $m \geq 2$), or comes from genuine cubic
interactions. These contributions are
therefore subleading.

The
fundamental mode therefore still obeys the linear equation
\begin{equation}
	\ddot j_1 + c_s^2 k^2 j_1 + \frac{\etaL}{w_0} k^2 \dot j_1 = 0.
	\label{eq:j1_linear}
\end{equation}
For \(\varepsilon\ll1\), its 
solution can be written as
\begin{equation}
	j_1(t) = J_1 e^{-\omega_I t}\cos(\omega_R t),\qquad
	\omega_R = c_s k,\qquad
	\omega_I = \frac{\etaL}{2w_0} k^2.
	\label{eq:j1_sol}
\end{equation}
The corresponding energy perturbation follows from the continuity equation
\(\dot e_1+k j_1=0\). 
Choosing the integration constant so that
the spatial average vanishes,
\begin{equation}
	e_1(t) = -\frac{J_1}{c_s} e^{-\omega_I t}\sin(\omega_R t), 
	\label{eq:e1_sol}
\end{equation}
up to corrections of order $\omega_I/\omega_R$ which is negligible in the long wavelength limit. 
Thus the fundamental mode is the slowest decaying mode and serves as the seed from which the higher-harmonic cascade is generated.

\section{Nonlinear equations of motion}
\label{sec:n3}

As discussed above, nonlinear terms must be kept in the
momentum sectors $\{nk,n>1\}$. 
We first expand the constitutive relation to the nonlinear power needed for the later projection, using the rest-frame
energy perturbation \(e=\epsilon-\epsilon_0\) and the velocity \(v\). Here the
power of a field should not be identified directly with its late-time
amplitude order; different momentum sectors can require different
truncations.
The
equation of state is
\be
	p=p_0+c_s^2 e+\frac12 p_0''e^2+\cdots,
	\qquad {\rm with~~}
	p_0''=\Bigl(\frac{\partial^2p}{\partial\epsilon^2}\Bigr)_0 .
\ee
Using the conventions defined in \eqref{eq:EJ_def_neutral},  
one obtains
\begin{align}
	\mathcal{E}
	&= \epsilon_0 + e + w_0 v^2 + \cdots,
	\label{E2}\\
	J
	&= w_0 v + (1+c_s^2)e v - \etaL\,v\partial_x v 
	+ \cdots,
	\label{J2}\\
	\Pi
	&= p_0 + c_s^2 e + \frac12 p_0''e^2 + w_0 v^2
	-\etaL\partial_x v
	+ \cdots.
	\label{Pi2}
\end{align}
The term ``\(\cdots\)" collects possible higher-gradient contributions and higher power terms.  Their 
explicit form is not needed for the analysis below.

Inverting \eqref{E2}--\eqref{J2} to the order needed here and using \eqref{eq:E_def_neutral} gives
\begin{equation}
	v = \frac{J}{w_0} - \frac{1+c_s^2}{w_0^2}EJ
	+\frac{\etaL}{w_0^3}J\partial_xJ+\cdots,
	\qquad
	e = E - \frac{J^2}{w_0}+\cdots .
	\label{eofEJ}
\end{equation}
Substituting these relations into \eqref{Pi2} gives
\begin{align}
	\Pi = p_0
	&+ c_s^2 E + \frac12 p_0''E^2
	+ \frac{1-c_s^2}{w_0}J^2
	- \frac{\etaL}{w_0}\partial_x J
	\notag\\
	& 
    + \partial^{1} N^{2+}+ N^{3+}+\partial^{2+}N^{1+}.
	\label{PiEJ}
\end{align}
The first line contains the ideal quadratic terms and the linear viscous term.
The second line collects three classes of nonlinear terms:  
first-order gradient nonlinear terms $\partial^{1} N^{2+}$, higher power terms $N^{3+}$, and higher-gradient expansion terms  $\partial^{2+} N^{1+}$. 
Here \(N\) denotes either \(E\) or
\(J\). The term 
\(N^{l+}\) denotes terms of power \(l\) or
higher in the fields $N$.  Similarly, \(\partial^q\) denotes terms with \(q\) derivatives, and
\(\partial^{q+}\) denotes terms with \(q\) or more derivatives.

The conservation equations become
\begin{align}
	\partial_tE+\partial_xJ&=0,
	\label{eq:neutral_energy_EJ}\\
	\partial_tJ+c_s^2\partial_xE
	-\frac{\etaL}{w_0}\partial_x^2J
	+\frac{2(1-c_s^2)}{w_0}J\partial_xJ
	+p_0''E\partial_xE
	&=-\partial_x\left(\partial^{1} N^{2+}+ N^{3+}+\partial^{2+}N^{1+}\right).
	\label{eq:neutral_momentum_EJ}
\end{align}
Combining the two equations gives
\begin{align}
	\partial_t^2J
	-c_s^2\partial_x^2J
	-\frac{\etaL}{w_0}\partial_x^2\partial_tJ
	&=
	-\frac{2(1-c_s^2)}{w_0}
	\partial_t(J\partial_xJ)
	-p_0''\partial_t(E\partial_xE)
	-\partial_t\partial_x\left(\partial^{1} N^{2+}+ N^{3+}+\partial^{2+}N^{1+}\right).
	\label{eq:J_nonlinear_wave_E_derivatives}
\end{align}

To obtain the equation of motion for each momentum, we  project \eqref{eq:J_nonlinear_wave_E_derivatives} onto the $nk$ sector, i.e. keep the terms with $\sin nkx$ spatial dependence. The term 
\(\partial_t\partial_x\left(\partial^{1} N^{2+}+ N^{3+}+\partial^{2+}N^{1+}\right)\)
generates different contributions in different momentum sectors. In the
long-wavelength regime,
\(\partial_t\partial_x\left(\partial^{1} N^{2+}\right)\) and
\(\partial_t\partial_x\left(\partial^{2+} N^{1+}\right)\) are suppressed by
extra powers of momentum. The \(N^{3+}\) terms, however, can contain different
structures with the same decay rate. Therefore, in each projected equation, we must identify and retain the dominant terms. We will do this explicitly below. 

\subsection{Second harmonic equation and its solution}
\label{sec:power_counting}
We first consider the equation of motion for $2k$ mode by projecting \eqref{eq:J_nonlinear_wave_E_derivatives} onto the
\(\sin(2kx)\) component. The term $J\partial_xJ$ contains contributions such as  
$j_1(t,x)\partial_x j_1(t,x)$, $j_{m+2}(t,x)\partial_x j_{m}(t,x)$ and $j_m(t,x)\partial_x j_{m+2}(t,x)$, etc. 
{From the hierarchy discussed around
\eqref{eq:nonlinear spectrum}, terms involving higher momenta are negligible at
late times. For the same reason, the \(N^{3+}\) terms do not contribute at the
leading order in the second-harmonic equation. 

After truncating all subleading 
terms, 
the equation of motion for $2k$ mode becomes 
\begin{align}
	\ddot j_2 + 4c_s^2 k^2 j_2
	+4\frac{\etaL}{w_0}k^2\dot j_2
	= S_2\,,\label{eq:2keq}
    \end{align}
    where $S_2$ is the nonlinear source term of the form 
 \begin{align}   
    S_2=-\frac{d}{dt}\left[\frac{(1-c_s^2)k}{w_0}j_1^2
	-\frac{p_0''k}{2}e_1^2 \right]. 
	\label{eq:j2_full}
\end{align}
 In this equation, we have factored out the common spatial dependence of $\sin 2kx$, and converted the spatial gradient into a factor of $k$.

Using \eqref{eq:j1_sol} and \eqref{eq:e1_sol}, the nonlinear source term on the right side becomes 
\begin{align}
	j_1^2
	&= \frac{J_1^2}{2}e^{-2\omega_I t}
	\bigl[1+\cos(2\omega_R t)\bigr],~~~~~~~
	e_1^2
	= \frac{J_1^2}{2c_s^2}e^{-2\omega_I t}
	\bigl[1-\cos(2\omega_R t)\bigr].
\end{align}
Therefore $S_2$ becomes
\begin{align}
	S_2
	&=-\frac{d}{dt}\left[ A_0 e^{-2\omega_I t}
	+A_2 e^{-2\omega_I t}\cos(2\omega_R t)\right]\\
    &=\,2\omega_I A_0 e^{-2\omega_I t}
	+2A_2\sqrt{\omega_I^2 + \omega_R^2} \, e^{-2\omega_I t} \sin\bigl(2\omega_R t + \arctan\frac{\omega_I}{\omega_R}\bigr)\\
    &\approx\,2\omega_I A_0 e^{-2\omega_I t}
	+2A_2\omega_R\, e^{-2\omega_I t} \sin\bigl(2\omega_R t\bigr),
    \end{align}
    with 
   \begin{align} 
	A_0
	\equiv
	\frac{kJ_1^2}{2w_0}
	\left(
	1-c_s^2-\frac{w_0p_0''}{2c_s^2}
	\right)=\frac{k\lambda J_1^2}{2w_0},~~~~~~~
	A_2
	\equiv
	\frac{kJ_1^2}{2w_0}
	\left(
	1-c_s^2+\frac{w_0p_0''}{2c_s^2}
	\right)
	=
	\frac{k\Lambda J_1^2}{2w_0},
\end{align}
where
\begin{equation}
\lambda\equiv1-c_s^2-\frac{w_0p_0''}{2c_s^2},\quad ~~~
	\Lambda
	\equiv
	1-c_s^2+\frac{w_0p_0''}{2c_s^2}.
	\label{eq:Lambda_def}
\end{equation}
Here $\lambda$ and $\Lambda$ are quantities determined by the thermodynamic properties of the system.

Thus the source separates into a non-resonant part with 
\(\omega=0\) and a resonant part with 
\(\omega=2\omega_R\). Here ``resonant" means that the source  frequency {matches the corresponding linear frequency.}

We write the particular solution directly in amplitude-phase form,
\begin{equation}
	j_2(t)
	=
	e^{-2\omega_I t}
	\Bigl[
	B+D\cos(2\omega_Rt-\phi_2)
	\Bigr].
	\label{eq:j2_phase_ansatz}
\end{equation} 
The $B$ term is the non-resonant response to the source, whereas the $D$ term is the resonant response. 
Substituting \eqref{eq:j2_phase_ansatz} into \eqref{eq:2keq} gives
\begin{equation}
	B
	\approx\frac{\omega_I A_0}{2\omega_R^2}, \quad ~~~
    D
	\approx
	\frac{|A_2|}{4\omega_I},
	\label{eq:B_exact}
\end{equation}
where we have used the relation between $\omega_R$, $\omega_I$ and $k$ in \eqref{eq:j1_sol}. The phase is
\begin{equation}
	\phi_2\approx
	\begin{cases}
	\displaystyle
	\pi,
	& \Lambda>0,\\[0.5em]
	\displaystyle
	0,
	& \Lambda<0.
	\end{cases}
	\label{eq:phi2_piecewise}
\end{equation}
For \(\Lambda=0\), the leading resonant quadratic source vanishes, and the $2k$ response 
becomes purely decaying at this order. 
This is an interesting special case.\footnote{This is reminiscent of the Bethe--Zel'dovich--Thompson (BZT) regime of
nonlinear acoustics, where the leading quadratic acoustic nonlinearity
vanishes.  The simple resonant quadratic cascade is then absent, and the
late-time attractor should be controlled by subleading non-resonant terms,
cubic nonlinearities, or finite-\(k\) corrections.}
In what 
follows we focus on the $\Lambda\neq0$ case and we restrict to the long-wavelength condition so that 
the non-resonant response is much smaller than the resonant response. {Since
\(B/D\sim 2(\omega_I/\omega_R)^2|\lambda/\Lambda|\), } this condition can be written as 
\begin{align}
    \frac{\omega_I}{\omega_R}
    =\frac{\eta_L k}{2 w_0 c_s} \ll 
\sqrt{\frac{|\Lambda|}{2|\lambda|}}. 
\label{eq:resonant condition}
\end{align}
{Under this condition, the leading \(2k\) solution is}
\begin{align}
\begin{split}
j_2(t)&=\frac{|A_2|}{4\omega_I}e^{-2\omega_It}\cos(2\omega_R t-\phi_2)
=\frac{|\Lambda|}{4\etaL k}J_1^2e^{-2\omega_It}\cos(2\omega_R t-\phi_2)\\
&=J_2e^{-2\omega_It}\cos(2\omega_R t-\phi_2),\\
	e_2(t)& =- \frac{J_2}{c_s} e^{-2\omega_It}\sin(2\omega_R t-\phi_2).
\end{split}
\end{align}

Equivalently, we define the second-harmonic cascade coefficient 
\begin{equation}
	\alpha_{J}^{(2)}\equiv\frac{J_2}{J_1^2}=\frac{|\Lambda|}{4\etaL k}.
	\label{eq:alphaJ_signed}
\end{equation}
{We will show below that, in the late-time and
long-wavelength regime, the same coefficient controls all \(nk\) harmonics. Before turning to general \(n\), we first study the \(3k\) mode, whose analysis
can be simplified by using the \(2k\) result.}

\subsection{Third harmonic equation and its solution}
For the $3k$ mode, we project the nonlinear terms onto the $\sin 3kx$ component and identify the dominant contribution at late times. 
{This
sector contains an additional complication from the \(N^{3+}\) terms: quadratic
sources and cubic sources can have the same decay rate and can therefore both
contribute to the \(3k\) source. Schematically,}
\begin{align}
j_1(t)j_2(t)\propto e^{-3\omega_I t}\,,~~~~~~~~~~~~
    j_1(t)^3\propto e^{-3\omega_I t}.
\end{align}

Thus, in a generic power expansion, 
we should keep the $N^3$ terms in solving the $3k$ equation. However, the analysis of the $2k$ momentum mode shows that the quadratic and cubic contributions are separated by their prefactors 
\begin{align}
j_1(t)j_2(t)\propto\frac{|\Lambda|}{4\eta_Lk}J_1^3 e^{-3\omega_I t}\,,~~~~~~~
    {\frac{j_1(t)^3}{w_0}\propto \frac{J_1^3}{w_0} e^{-3\omega_I t},}
\end{align}
The $\frac{1}{w_0}$ factor in the $j_1(t)^3$ term is the factor in equations of motion that matches the dimensions (assuming that the relevant dimensionless thermodynamic derivatives remain finite and of order unity). Therefore, if we restrict to the long-wavelength condition 
\begin{align}
    {w_0 \alpha_{J}^{(2)}}=\frac{|\Lambda| w_0}{4\eta_Lk}\gg 1, 
\end{align}
{the quadratic source \(j_1j_2\) dominates over the cubic
source \(j_1^3\). } 
Under this condition, the cubic terms can be neglected in the leading equations for $3k$, which becomes
\begin{align}
    \ddot j_3+9\omega_R^2j_3+18\omega_I\dot j_3
	=S_3\,,
    \end{align}
where the source 
    \begin{align}
    S_3=-\frac{d}{dt}\left[\frac{3(1-c_s^2)k}{w_0}j_1j_2
	-\frac{3p_0''k}{2}e_1 e_2 \right].
\end{align}

Following the same procedure as in the $2k$ calculation, one obtains 
\begin{align}
    j_3(t)&=J_3e^{-3\omega_I t}\cos(3\omega_Rt-\phi_3),
~~~~~~  e_3(t)=- \frac{J_3}{c_s} e^{-3\omega_It}\sin(3\omega_R t-\phi_3).   
    \end{align}
    where 
\begin{align}    
    J_3=\frac{|\Lambda|}{4\etaL k}J_1J_2,~~~~~~
    \phi_3=0.
\end{align} 
In this \(3k\) analysis, the non-resonant part is also
negligible compared to the resonant part under the corresponding long-wavelength condition for this harmonic, which is 
\begin{align}
    \frac{\omega_I}{\omega_R}
    =\frac{\eta_L k}{2 w_0 c_s} \ll {\frac{|\Lambda|}{|\lambda|}}.
\end{align}
This gives a different constraint from \eqref{eq:resonant condition} in the
\(2k\) analysis, 
because the non-resonant source in the \(3k\) sector carries a
nonzero real frequency, unlike the zero-frequency non-resonant component in the
\(2k\) case. 

We therefore obtain 
\begin{align}
    \alpha_{J}^{(3)}\equiv \frac{J_3}{J_1J_2}=\frac{|\Lambda|}{4\eta_L k}.
\end{align}
Thus the third harmonic exhibits the same cascade coefficient that appeared in the second-harmonic solution \eqref{eq:alphaJ_signed}. From now on, we omit the harmonic label on the cascade
coefficient and define 
\begin{align}
    \alpha_J=\frac{|\Lambda|}{4\eta_L k}.
\end{align}

\subsection{General harmonic equation and its solution}
{Guided by the \(2k\) and \(3k\) results, we now derive the late-time solution for a general \(nk\) mode. We work under the
long-wavelength and dominance conditions introduced above, namely} 
\begin{align}
    \frac{\omega_I}{\omega_R}\ll1\quad \text{and}\quad {w_0\alpha_J}\gg1. 
\end{align}

Assume that all harmonics below \(n\) take the locked late-time form
\begin{equation}
	j_m(t)
	=J_me^{-m\omega_I t}\cos(m\omega_Rt-\phi_m),
	\qquad J_m>0,
	\qquad {\rm for~any~} m<n,
	\label{eq:jn_phase_ansatz}
\end{equation}
with the corresponding energy perturbation 
\begin{equation}
	e_m(t)
	= -\frac{J_m}{c_s}e^{-m\omega_I t}
	\sin(m\omega_Rt-\phi_m).
	\label{eq:en_phase}
\end{equation}
The amplitude hypothesis is
\begin{equation}
	J_m=\alpha_J^{m-1}J_1^m,
	\qquad
	\alpha_J=\frac{|\Lambda|}{4\etaL k}.
	\label{eq:amplitude_cascade}
\end{equation}
and the phase hypothesis {follows the pattern found for the
second and third harmonics,
\begin{align}
    {
    \phi_m=
	\begin{cases}
	(m-1)\pi, & \Lambda>0,\\
	0, & \Lambda<0,
	\end{cases}
	\qquad \mathrm{mod}\ 2\pi .}
\label{eq:phase-ass}
\end{align}
}

The leading equation of motion for the $nk$ momentum mode is
\begin{align}
	&\ddot j_n+n^2(c_sk)^2j_n+2n^2\omega_I\dot j_n
	=S_n 
    \label{eq:jn_full} 
    \end{align}
 with the source term   
 \begin{align}
    &S_n=-\frac{d}{dt}\left[\frac{(1-c_s^2)k}{w_0}
	\sum_{m=1}^{n-1}(n-m)j_mj_{n-m}
	-\frac{p_0''k}{2}
	\sum_{m=1}^{n-1}(n-m)e_m e_{n-m}\right],
\end{align}
where, as in the $2k$ and $3k$ mode, we keep the leading
quadratic source and neglect the \(N^{3+}\) terms and
non-resonant pieces under similar long-wavelength 
conditions, specifically 
\begin{align}
    \frac{\omega_I}{\omega_R}
    =\frac{\eta_L k}{2 w_0 c_s} \ll {\frac{|\Lambda|}{n|\lambda|}}.
\end{align}
This condition determines how large $n$ can be. Under this condition, we obtain the solution for the $nk$ mode
\begin{align}
    j_n(t)=J_ne^{-n\omega_I t}\cos(n\omega_Rt-\phi_n),~~~~~
    e_n(t)
	= -\frac{J_n}{c_s}e^{-n\omega_I t}
	\sin(n\omega_Rt-\phi_n). 
\end{align}
with
\begin{align}
    J_n=\frac{|\Lambda|}{2n(n-1)\etaL k}
	\sum_{m=1}^{n-1}(n-m)J_mJ_{n-m}=\alpha_JJ_1J_{n-1}
\end{align}
and the phase $\phi_n$ is  
\begin{align}
    \phi_n=&
	\begin{cases}
	\displaystyle
	(n-1)\pi,
	& \Lambda>0,\\[0.6em]
	\displaystyle
	0,
	& \Lambda<0,
	\end{cases}
    \qquad \mathrm{mod}\ 2\pi .
	\label{eq:phin_piecewise}
\end{align}
The resulting locked tower gives the late-time attractor
solution for the viscous hydrodynamic system within the controlled
long-wavelength regime. 
More precisely, to have such a simple cascade structure, $nk$ mode should satisfy 
\begin{align}
    \frac{n\omega_I}{\omega_R}\ll1, \quad \frac{\omega_I}{\omega_R}
    =\frac{\eta_L k}{2 w_0 c_s} \ll {\frac{|\Lambda|}{n|\lambda|}} \quad\text{and}\quad {w_0\alpha_J}\gg1,\quad \text{for}\ \ n\ge3\,.
    \label{eq:regime}
\end{align}
Here we impose the weak damping condition to the $nk$ mode (the first condition). Together with
the resonant-dominance condition (the second condition) and the quadratic-source dominant condition (the third condition), these
define the regime in which the simple formula derived above is valid.
Conversely, once the fundamental momentum $k$ is fixed, they
determine the range of harmonics for which the simple attractor expression
applies.

For harmonics outside this controlled range, the late-time behavior may still be governed by nonlinear sources determined by the fundamental mode rather than by independent initial data, but the frequency, phase shift, and cascade coefficient \(\alpha_J\) will acquire \(k\)-dependent corrections.  The simple cascade shown before terminates when the intrinsic linear decay of a harmonic becomes slower than all available nonlinear source
terms.  The precise endpoint therefore depends on the full dispersion relation of the system.

\section{Late-time attractor and limiting cases}
\label{sec:cascade_conclusion}

In the long-wavelength regime, we obtain a simple solution describing the relaxation of a near-equilibrium dissipative hydrodynamic system. The attractor {is characterized by frequency locking,}
$\omega_n\rightarrow n(\omega_R-i\omega_I)$, and by the cascade coefficient $\alpha_J=\frac{|\Lambda|}{4\eta_L k}$ where \(\Lambda\) is fixed by the 
thermodynamic properties of the fluid. Here we can examine several
representative limits of the attractor.

\subsection{Conformal limit}

For a neutral conformal fluid in \(d+1\) spacetime dimensions,
\be
	p=\frac{\epsilon}{d},\qquad
	c_s^2=\frac1d,\qquad
	p_0''=0,\qquad
	\zeta=0 .
\ee
Then
\be
	\Lambda=1-\frac1d=\frac{d-1}{d},
	\qquad
	\etaL=\frac{2(d-1)}{d}\eta .
\ee
The cascade coefficient therefore reduces to a $d$-independent quantity 
\begin{equation}
	\alpha_J^{\rm conf}
	=\frac{1}{8\eta k}.
	\label{eq:alphaJ_conf}
\end{equation} 
This provides a
useful check of the normalization and agrees with the coefficient extracted in
the conformal holographic setup of Ref.~\cite{Liu:2025hfv}.

Since \(\alpha_J\) has dimensions inverse to momentum density, it is useful to
form a dimensionless measure using the entropy density \(s\) and the period
\(L=2\pi/k\):\footnote{Here the subscript $d$ labels the spatial dimension of the fluid.}
\begin{equation}
	\bar \alpha_{J_{(d)}}
	\equiv \alpha_J\frac{s}{L}.
	\label{eq:alphabar_def_EJ}
\end{equation}
For a neutral conformal fluid this gives
\begin{equation}
	\bar\alpha_{J_{(d)}}^{\rm conf}
	= \frac{s}{16\pi\eta}.
	\label{eq:alphabar_neutral_EJ}
\end{equation}
If the fluid satisfies the KSS bound \(\eta/s\ge 1/(4\pi)\) \cite{Kovtun:2004de}, then
\begin{equation}
	\bar\alpha_{J_{(d)}}^{\rm conf}\le\frac14 .
	\label{eq:alphabar_conformal_eta_s_EJ}
\end{equation}
Thus the conformal result gives a simple dimensionless upper estimate for the
strength of the neutral fluid cascade.

\subsection{Non-relativistic compressible fluids}

The key ingredients of the late-time attractor are the $k^{2}$ sound attenuation and 
the quadratic nonlinearities of the conservation laws.  Neither of them relies on 
the relativistic constitutive relation. It is
therefore natural to expect the same mechanism to appear in a
non-relativistic compressible fluid, if 
the system is effectively isentropic and sound remains the only long-lived mode.

Consider a one-dimensional fluid without heat conduction, described by the mass 
density $\rho(t,x)$, velocity $u(t,x)$ and a barotropic equation of state 
$p=p(\rho)$.  The mass and momentum conservation equations are \cite{Landau:1987}
\begin{equation}
  \partial_{t}\rho+\partial_{x}(\rho u)=0,\qquad
  \partial_{t}(\rho u)+\partial_{x}(\rho u^{2}+p)
  = \partial_{x}(\mu\,\partial_{x}u),
\end{equation}
where $\mu\equiv \zeta +2\frac{d-1}{d}\eta$ denotes the longitudinal viscosity.  Introducing the mass-density 
perturbation $\delta\rho\equiv\rho-\rho_{0}$ and the momentum density 
$j\equiv\rho u$, and expanding the pressure to second order,
\be
p = p_{0} + c_{s}^{2}\,\delta\rho + \frac12 p_{0}''(\delta\rho)^{2} 
+ \cdots,\qquad
c_{s}^{2} = \Bigl(\frac{dp}{d\rho}\Bigr)_{0},\quad
p_{0}'' = \Bigl(\frac{d^{2}p}{d\rho^{2}}\Bigr)_{0},
\ee
one obtains, after eliminating $\delta\rho$ and keeping quadratic terms, a 
nonlinear wave equation structurally identical to \eqref{eq:J_nonlinear_wave_E_derivatives}:
\begin{equation}
  \partial_{t}^{2}j - c_{s}^{2}\partial_{x}^{2}j 
  - \frac{\mu}{\rho_{0}}\partial_{x}^{2}\partial_{t}j
  = -\frac{2}{\rho_{0}}\,\partial_{t}(j\partial_{x}j)
  - p_{0}''\,\partial_{t}(\delta\rho\,\partial_{x}\delta\rho).
  \label{eq:nr_wave}
\end{equation}
This matches the relativistic equation under the replacement
\begin{equation}
  w_{0}\rightarrow\rho_{0},\qquad
  \eta_{L}\rightarrow\mu,\qquad
  E\rightarrow\delta\rho,\qquad
  J\rightarrow j,
  \label{eq:nr_map}
\end{equation}
with the thermodynamic coefficient $\Lambda_{\rm nr}$ becoming\footnote{It is interesting to note that this coefficient matches precisely with the fundamental derivative in non-relativistic fluid. }
\begin{equation}
  \Lambda_{\rm nr}=1+\frac{\rho_{0}\,p_{0}''}{2c_{s}^{2}}.
  \label{eq:Lambda_nr}
\end{equation}

For the fundamental mode, the linear part of \eqref{eq:nr_wave} reduces to the 
damped harmonic oscillator, giving
\begin{equation}
  j_{1}(t) = J_{1} e^{-\omega_{I}^{\rm nr} t}\cos(\omega^{\rm nr}_{R}t),\qquad
  \delta\rho_{1}(t) = -\frac{J_{1}}{c_{s}} e^{-\omega_{I}^{\rm nr} t}\sin(\omega^{\rm nr}_{R}t),
  \label{eq:nr_fundamental}
\end{equation}
with $\omega_{R}^{\rm nr}=c_{s}k$ and $\omega_{I}^{\rm nr}=\mu k^{2}/(2\rho_{0})$.

The three long-wavelength conditions that guarantee the simple cascade carry
over directly. 
One requires
\begin{align}
  \frac{n\mu k}{2\rho_{0}c_{s}}\ll 1,\qquad
  \frac{\mu k}{2\rho_{0}c_{s}}\ll \frac{|\Lambda_{\rm nr}|}{n|\lambda_{\rm nr}|},\qquad
  \frac{\rho_{0}|\Lambda_{\rm nr}|}{4\mu k}\gg 1,
  \label{eq:nr_conditions}
\end{align}
where $\lambda_{\rm nr}=1-\rho_{0}p_{0}''/(2c_{s}^{2})$.  The first inequality 
is the weak-damping condition, the second suppresses the non-resonant driven 
response, and the third guarantees that quadratic sources dominate. 

Under the conditions \eqref{eq:nr_conditions} the derivation proceeds exactly 
as in the relativistic case, yielding the same attractor
\begin{align}
    j(t)=\sum_{n}J^{\rm nr}_n e^{-n\omega^{\rm nr}_I t}\cos (n\omega^{\rm nr}_R t-\phi_n), 
\end{align}
frequency locking 
$\omega_{n}^{\rm nr}\rightarrow n(\omega_{R}^{\rm nr}-i\omega_{I}^{\rm nr})$, $\phi_n$ takes the same form as \eqref{eq:phin_piecewise}, and amplitude cascade 
$J^{\rm nr}_{n}=(\alpha_{J}^{\rm nr})^{n-1}(J_{1}^{\rm nr} )^{\,n}$, with\footnote{It is interesting to note that this result can be obtained from the infinite speed of light limit  $c\to\infty$ of relativistic result \eqref{eq:Lambda_def}.  
 Restoring \(c\) in \eqref{eq:Lambda_def}, take the limit $	c_s\ll c,
	p_0\ll \epsilon_0,
	w_0\simeq \rho_0c^2,$ one recovers $\Lambda_\text{nr}$ in the non-relativistic fluid.   
}
\begin{equation}
  \alpha_{J}^{\rm nr}  = \frac{|\Lambda_{\rm nr}|}{4\mu k}.
  \label{eq:alphaJ_nr}
\end{equation}
For a polytropic gas $p\propto\rho^{\gamma}$ one has 
$c_{s}^{2}=\gamma p_{0}/\rho_{0}$, $p_{0}''=\gamma(\gamma-1)p_{0}/\rho_{0}^{2}$, 
so that $\Lambda_{\rm nr}=1+\frac{\gamma-1}{2}=\frac{\gamma+1}{2}$ and 
$\alpha_{J}^{\rm nr}=(\gamma+1)/(8\mu k)$.  For a monatomic gas 
$\gamma=5/3$ this gives $\alpha_{J}^{\rm nr}=1/(3\mu k)$. It would be interesting to test the cascading relations in these real systems.  

The complete structural equivalence under the map \eqref{eq:nr_map} shows that 
the late-time hydrodynamic attractor is a robust consequence of the 
one-dimensional conservation laws with $k^{2}$ sound attenuation, independent of 
the microscopic origin of the fluid and of relativistic covariance.


\section{Discussion and Conclusion}
\label{sec:discussion}

We have shown that the late-time relaxation of a near-equilibrium viscous fluid
in a compact space is governed by a nonlinear attractor, not by a superposition
of independent linear quasinormal modes.  The attractor is characterized by three
quantitative features: (i) frequency locking,
\(\omega_n\to n(\omega_R-i\omega_I)\); (ii) an amplitude cascade,
\(J_n=\alpha_J^{\,n-1}J_1^n\), with the cascade coefficient
\(\alpha_J=|\Lambda|/(4\eta_L k)\) determined entirely by equilibrium
thermodynamics and the longitudinal viscosity; and (iii)  phase locking, 
when \(\Lambda>0\), \(\phi_n=(n-1)\pi\); when \(\Lambda<0\), \(\phi_n=0\); there is no free phase slippage between harmonics.

The origin of this behavior is the hierarchy already present in the linear
spectrum: \(\omega_{nI}=n^2\omega_I\) implies that higher Fourier modes are
exponentially more damped than lower ones.  Nonlinear products of
slower-decaying modes therefore act as sources for higher-momentum sectors; these
sources generically decay more slowly than the intrinsic linear mode of the
target sector.  For example, the quadratic product \(j_1^2\) decays as
\(e^{-2\omega_I t}\) while the linear \(2k\) mode decays as \(e^{-4\omega_I
t}\).  At late times the nonlinear source inevitably dominates, regardless
of how small the initial perturbation amplitude may be.

A crucial methodological point is that the power of fluctuation fields in the
equations of motion is not equivalent to their late-time amplitude order. 
A consistent truncation therefore requires a joint power-momentum analysis:
one must retain all terms that can contribute at leading order in each
momentum sector separately.

The simple cascade with constant \(\alpha_J\) operates within a controlled
long-wavelength regime defined by three conditions \eqref{eq:regime} on the \(n\)-th harmonic. Meanwhile, 
the existence of the attractor relies on discrete momenta \(k_n=nk\).  In the
continuum limit \(L\to\infty\),
the same nonlinear effect should persist. 
The loss of sharp harmonic separation turns
nonlinear products into momentum-space convolutions that depend on the global
shape of the initial spectrum, rather than on a single fundamental-mode
amplitude.

For conformal fluids the dimensionless cascade strength is
\(\bar\alpha_J^{\rm conf}=s/(16\pi\eta)\).  Together with the KSS bound~\cite{Kovtun:2004de}, this yields the universal upper
bound
$
  \bar\alpha_J^{\rm conf}\le\frac14 .
$
The bound is saturated by strongly coupled fluids near the minimal viscosity.
This establishes a quantitative link between the strength of the nonlinear
harmonic cascade and the uncertainty principle as manifested in the
shear viscosity bound.  Weakly coupled fluids (large \(\eta/s\)) exhibit
suppressed cascades, while holographic quantum critical systems and the
quark-gluon plasma lie near the maximal cascade regime.

The same attractor structure emerges in non-relativistic compressible fluids
with \(k^2\) sound attenuation, where \(\Lambda_{\rm nr}\)
coincides with the fundamental derivative of gas dynamics
\cite{Thompson1971,Landau:1987}, the dimensionless parameter that controls
nonlinear sound-wave steepening.  It is interesting to note that in the inviscid limit \(\mu\to0\),
the cascade coefficient diverges, the multi-harmonic shock precursor
re-emerges, and the attractor can be viewed as the dissipative
regularization of nonlinear steepening.

There are several further directions for future study. 
First, a natural
next step is to compute the coefficients for 
charged fluids and holographic systems \cite{inpro, Liu:2025hfv} and discuss the nonlinear solutions in holography \cite{Liu:2025hfv, Pan:2024bon, Emparan:2026lss}. 
Second, it would also be useful to study  hydrodynamic systems with diffusion-to-sound crossover \cite{Grozdanov:2018fic}. Third, it would be interesting to study the nonlinear harmonics in causal first-order hydrodynamics \cite{Kovtun:2019hdm, Bemfica:2019knx}, fracton hydrodynamics \cite{Glodkowski:2022xje}, non-boost invariant  hydrodynamics \cite{deBoer:2020xlc} and hydrodynamics of charged density wave phases \cite{Baggioli:2022pyb}. Finally, systems near criticality, where bulk viscosity and susceptibilities can become large, may provide a sensitive arena for testing the robustness of the nonlinear harmonic cascade. It would also be important to identify experimental setups and observables that could directly test the predicted harmonic hierarchy.

\subsection*{Acknowledgments} 
We thank Yan-Yan Bu, Elias Kiritsis, Wen-Bin Pan, Ya-Wen Sun and Xin-Meng Wu for useful discussions. This work was supported by the National Natural Science Foundation of China Grants No. 12375041, 12447169 and 12575046. H-T. S also acknowledges the support from the Postdoctoral Fellowship Program of CPSF under Grant No. GZC20252777.

\end{document}